\documentclass[pdflatex,sn-mathphys-num]{sn-jnl}

\usepackage{graphicx}%
\usepackage{multirow}%
\usepackage{amsmath,amssymb,amsfonts}%
\usepackage{amsthm}%
\usepackage[title]{appendix}%
\usepackage{xcolor}%
\usepackage{textcomp}%
\usepackage{manyfoot}%
\usepackage{booktabs}%
\usepackage{algorithm}%
\usepackage{algorithmicx}%
\usepackage{algpseudocode}%
\usepackage{listings}%
\usepackage{siunitx}
\usepackage{tabularx}  
\usepackage{graphicx}
\usepackage{booktabs}
\usepackage{changepage}
\usepackage{array}
\usepackage{graphicx}
\usepackage{float}
\usepackage[export]{adjustbox} 

\newcommand{\authorbio}[3]{%
  \vspace{1em}%
  \noindent
  \begin{minipage}[t]{35mm} 
    \includegraphics[width=35mm,height=45mm,keepaspectratio,valign=t]{#1} 
  \end{minipage}%
  \hspace{1em}
  \begin{minipage}[t]{\dimexpr\textwidth-35mm-1em} 
    \small
    \textbf{#2} #3
  \end{minipage}%
  \par\vspace{1em}%
}

\theoremstyle{thmstyleone}%
\theoremstyle{thmstyletwo}%

\theoremstyle{thmstylethree}%

\begin{document}

\title[Article Title]{A Novel Hybrid Approach for Retinal Vessel Segmentation with Dynamic Long-Range Dependency and Multi-Scale Retinal Edge Fusion Enhancement}


\author[1]{\fnm{Yihao} \sur{Ouyang}}\email{yihaoouyang@163.com}
\equalcont{These authors contributed equally to this work.}

\author[1]{\fnm{Xunheng} \sur{Kuang}}\email{xunhengkuang@gmail.com}
\equalcont{These authors contributed equally to this work.}

\author[1]{\fnm{Mengjia} \sur{Xiong}}\email{mengjia\_xiong@163.com}

\author[2]{\fnm{Zhida} \sur{Wang}}\email{wangzhida888@126.com}

\author*[1]{\fnm{Yuanquan} \sur{Wang}}\email{wangyuanquan@scse.hebut.edu.cn}

\affil[1]{\orgdiv{School of Artificial Intelligence}, \orgname{Hebei University of Technology}, \orgaddress{ \city{Tianjin}, \postcode{300401}, \country{China}}}

\affil[2]{\orgdiv{NHC Key Lab of Hormones and Development and Tianjin Key Lab of Metabolic Diseases}, \orgname{Tianjin Medical University Chu Hsien-I Memorial Hospital \& Institute of Endocrinology}, \orgaddress{\city{Tianjin}, \postcode{300134}, \country{China}}}


\abstract{Accurate retinal vessel segmentation provides essential structural information for ophthalmic image analysis. However, existing methods struggle with challenges such as multi-scale vessel variability, complex curvatures, and ambiguous boundaries. While Convolutional Neural Networks (CNNs), Transformer-based models and Mamba-based architectures have advanced the field, they often suffer from vascular discontinuities or edge feature ambiguity. To address these limitations, we propose a novel hybrid framework that synergistically integrates CNNs and Mamba for high-precision retinal vessel segmentation. Our approach introduces three key innovations: 1) The proposed High-Resolution Edge Fuse Network is a high-resolution preserving hybrid segmentation framework that combines a multi-scale backbone with the Multi-scale Retina Edge Fusion (MREF) module to enhance edge features, ensuring accurate and robust vessel segmentation. 2) The Dynamic Snake Visual State Space block combines Dynamic Snake Convolution with Mamba to adaptively capture vessel curvature details and long-range dependencies. An improved eight-directional 2D Snake-Selective Scan mechanism and a dynamic weighting strategy enhance the perception of complex vascular topologies. 3) The MREF module enhances boundary precision through multi-scale edge feature aggregation, suppressing noise while emphasizing critical vessel structures across scales. Experiments on three public datasets demonstrate that our method achieves state-of-the-art performance, particularly in maintaining vascular continuity and effectively segmenting vessels in low-contrast regions. This work provides a robust method for clinical applications requiring accurate retinal vessel analysis. The code is available at \href{https://github.com/frank-oy/HREFNet}{https://github.com/frank-oy/HREFNet}.}




\maketitle

\section{Introduction}\label{sec1}
Accurate retinal vessel segmentation is a key task in medical image analysis. The morphological characteristics of retinal vessels are closely associated with various ophthalmic diseases, such as diabetic retinopathy, glaucoma, and hypertension \cite{imran2019comparative, zhang2018deep}.  
Therefore, accurate retinal vessel segmentation provides essential foundational data for early screening, disease monitoring, and subsequent research, thereby indirectly facilitating disease diagnosis and intervention \cite{li2023retinal}. Additionally, segmentation results help radiologists analyze vascular structures more intuitively, enabling better disease assessment and treatment planning \cite{fan2001enas}. With advancements in medical imaging, deep learning-based segmentation methods \cite{li2022novel} have shown great potential for efficient and high-precision segmentation of medical structures. However, challenges remain in achieving the same level of accuracy and robustness for vessel segmentation, particularly in complex or low-contrast regions.

Among existing methods, Convolutional Neural Networks (CNNs) have been widely adopted for retinal vessel segmentation due to their strong feature extraction capabilities. U-Net \cite{ronneberger2015u} and its variants \cite{zhou2018unet++, isensee2018nnu, li2018h} leverage encoder-decoder architectures, but are limited in modeling long-range dependencies and adapting to varying vessel structures. To address this, Transformer-based architectures have been introduced, such as TransUNet \cite{chen2021transunet}, which integrates Transformer modules into the U-Net backbone, and Swin-Unet \cite{cao2022swin}, which employs a hierarchical Transformer design to capture both global and local context. More recently, Mamba-based models have been explored for their efficient sequence modeling via state-space mechanisms. VM-UNet \cite{ruan2024vm} incorporates Mamba blocks into U-Net, improving segmentation of complex vessel structures, while Swin-UMamba \cite{liu2024swin} further enhances feature representation by combining Mamba with hierarchical Transformers. However, most existing Mamba-based models are not specifically designed for retinal vessels, which may limit their accuracy in capturing detailed structural information, especially for thin and intricate vessels. These limitations underscore the need for improved architectures that better capture multi-scale features and structural continuity in retinal vessel segmentation.

\begin{figure}[t!]
\centering
\includegraphics[width=0.95\columnwidth,keepaspectratio]{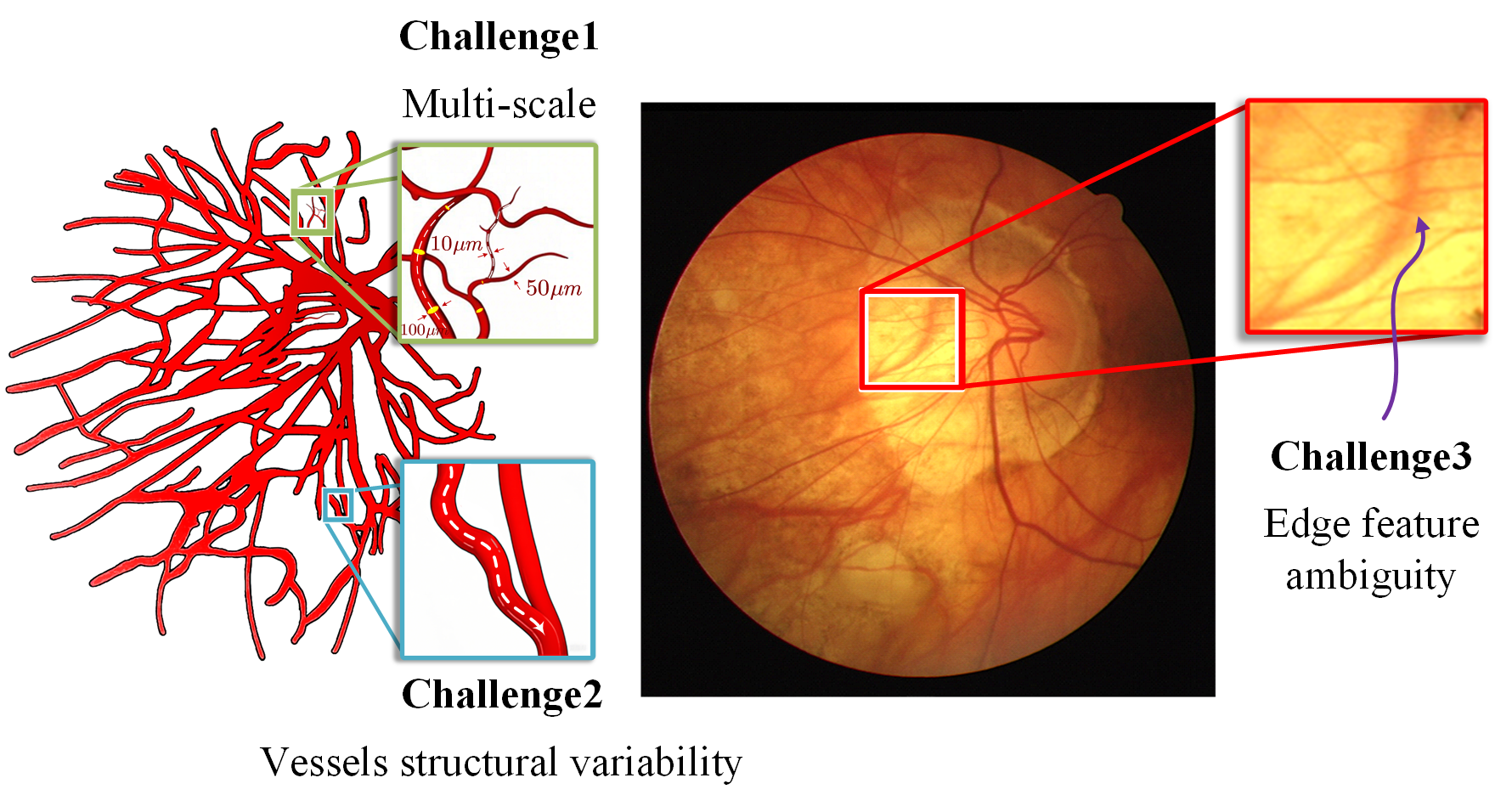}
\caption{\textbf{Challenges.} (1) Multi-scale. Vessels vary in size from large to small, requiring accurate multi-scale representation. (2) Vessels structural variability. Vessels exhibit diverse structural characteristics, including variations in both morphology and orientation. (3) Edge Feature Ambiguity. Low contrast and unclear boundaries make accurate segmentation difficult}
\label{fig:fig1}
\end{figure}

Although existing methods have made considerable progress, retinal vessel segmentation still faces several challenges. As illustrated in Fig. \ref{fig:fig1}, segmentation performance is significantly influenced by the following factors: \textbf{(1) Multi-scale}. Retinal vessels vary in scale, from thick primary vessels to thin capillaries. Effectively segmenting vessels across these scales remains challenging, as conventional methods often struggle to balance detail preservation with global structural consistency, leading to fragmented or incomplete vessels representations. \textbf{(2) Vessels structural variability}. Retinal vessels exhibit diverse structural characteristics, including variations in thickness, direction, and connectivity. Such complexity poses challenges for segmentation models, often resulting in discontinuous or distorted vessel representations, especially in highly curved or branched regions. \textbf{(3) Edge feature ambiguity}. The boundaries of retinal vessels are often indistinct due to low contrast and noise, making accurate edge delineation difficult. Additionally, weak texture information in fundus images further exacerbates the challenge of extracting clear vessel structures.
\begin{figure}[t!]
\centering
\includegraphics[width=0.95\columnwidth,keepaspectratio]{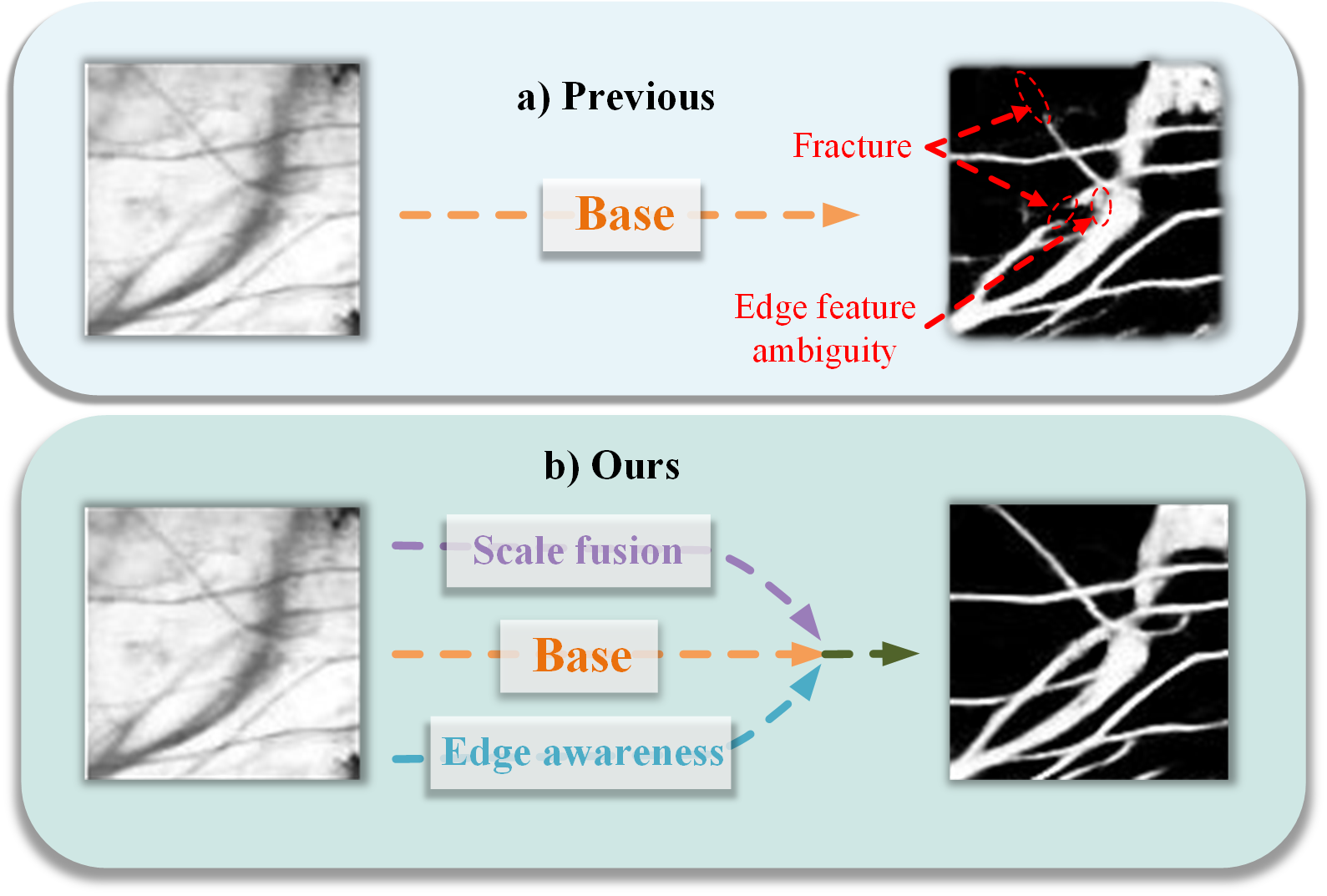}
\caption{\textbf{Motivation.} (a) The previous method relies solely on the base segmentation network, leading to edge feature ambiguity and vessel fracture (highlighted in \textcolor{red}{red}). (b) Our approach integrates scale fusion and edge awareness mechanisms, enhancing vessel continuity and reducing segmentation errors, particularly along thin and low-contrast vessels}
\label{fig:fig2}
\end{figure}

To address these challenges, we propose a hybrid segmentation framework integrating multi-scale segmentation and edge enhancement mechanisms, specifically optimized for the complex morphology of retinal vessels. As shown in Fig. \ref{fig:fig2}, our approach combines multi-scale fusion with edge awareness mechanisms, achieving edge awareness at different scales. The method is built upon a high-resolution preserving backbone, with an embedded multi-scale retinal edge fusion module that integrates multi-level edge features, effectively enhancing vessel boundary information and improving segmentation stability in low-contrast regions.
Firstly, our backbone network adopts HRNet \cite{sun2019deep}
, which preserves high-resolution representations throughout the process and enables continuous extraction of fine-grained features at multiple scales. This enhances the ability of the model to handle both large and small vessels, improving its perception of high-frequency information. As a result, the model achieves greater robustness in multi-scale vessel segmentation tasks.
Secondly, inspired by Dynamic Snake Convolution \cite{qi2023dynamic}, we propose the Dynamic Snake Visual State Space (DSVSS) module. Compared to the original method, we introduce a state-space modeling mechanism, enabling more accurate capture of long-range dependencies in vessels while enhancing adaptability to local curvature variations. This improvement allows for better segmentation of complex and curved vascular structures. Additionally, we adopt the 2D Selective Scan (SS2D) scanning mechanism \cite{liu2024vmamba} and extend it to eight directions to better capture complex vascular structures, particularly bifurcations and small vessels. Importantly, we introduce a dynamic weighting mechanism to adaptively aggregate multi-directional information, enhancing the accuracy of curvilinear structure tracing. Overall, this module strengthens local feature representation while improving long-range connectivity, enabling more complete vessel segmentation.
Finally, to address the issue of blurred vessel boundaries, we propose the Multi-scale Retina Edge Fusion (MREF) module, which enhances the preservation of vessel boundary information through a multi-scale edge feature fusion strategy. MREF aggregates edge features at different scales, allowing it to adaptively highlight key vessel boundaries while reducing background noise interference. Additionally, we introduce an attention mechanism to dynamically assign feature weights, enabling the model to more effectively focus on local vessel structures while maintaining overall boundary continuity. The introduction of this module further improves segmentation performance in low-contrast regions, thereby enhancing the overall accuracy and stability of vessel segmentation. The main contributions of this work are as follows:
\begin{itemize}
    \item \textbf{Hybrid vessel segmentation framework}: We propose High-Resolution Edge Fuse Network (HREFNet), a hybrid segmentation framework that integrates multi-scale segmentation and edge enhancement, built upon a high-resolution preserving backbone network. HREFNet combines the ability of the high-resolution preserving backbone to perform multi-scale fusion while introducing the Multi-scale Retina Edge Fusion module for edge perception, ensuring the accuracy, completeness, and robustness of vessel segmentation.
    \item \textbf{Dynamic Snake Visual State Space (DSVSS) block}: We propose the DSVSS block, which utilizes DSConv to adaptively match vessel curvature and integrates local feature extraction through CNNs with long-range modeling using Mamba to enhance small vessel detection and overall continuity. Additionally, the SS2D eight-directional scanning and dynamic weighting mechanism are improved to better adapt to complex vascular structures.
    \item \textbf{Multi-scale Retina Edge Fusion (MREF) module}: We introduce the MREF module, which enhances vessel boundary information through multi-scale feature fusion. It highlights key vessel edges at different scales, suppresses background noise, and improves segmentation stability in low-contrast regions.
    \item \textbf{Better performance on multiple datasets}: Our model achieves better results on three publicly available datasets, demonstrating its robustness and effectiveness in vessel segmentation across different types of retinal images.
\end{itemize}

\section{RELATED WORK}
\subsection{RETINAL VESSEL SEGMENTATION}
Retinal vessel segmentation has been extensively studied over the years, with traditional methods primarily relying on handcrafted features and classical image processing techniques. Early approaches often utilized  matched filtering, and morphological operations to enhance and extract vessel structures. For instance, Chaudhuri et al. \cite{chaudhuri1989detection} employed matched filtering to extract vessel features and estimate vessel diameter, using kernels across multiple scales and directions. While representative, the method is limited by its reliance on handcrafted design, making it less adaptable to complex conditions. In addition Frangi et al. \cite{frangi1998multiscale} proposed a vessel enhancement filter based on Hessian eigenvalues to enhance tubular structures, while Zhang et al. \cite{zhang2015retinal} employed graph-based algorithms to model vascular structures more effectively. However, these methods are highly sensitive to noise, illumination variations, and pathological artifacts, making them less robust in real-world clinical applications.

With the development of deep learning, many methods have been proposed that outperform traditional approaches. Dey et al. \cite{dey2020subpixel} proposed SpruNet, a subpixel convolution-based residual U-Net architecture that repurposes subpixel convolutions for down-sampling and up-sampling operations. The  subpixel convolution-based encoding-decoding strategy effectively reduces information loss, thereby increasing the sensitivity of the model without compromising specificity. Wu et al. \cite{wu2021scs} proposed SCS-Net, which dynamically adjusts the receptive field to extract multi-scale features and optimizes the fusion of adjacent hierarchical features to enhance vascular semantic information extraction, thereby effectively addressing challenges posed by vessel scale variations and complex backgrounds. In recent years, there has been an increasing focus on the topological and connectivity features of vascular structures \cite{liu2023wave, gao2020effective, lan2020elastic}. Wang et al. \cite{wang2020rvseg} proposed a feature pyramid cascade module that can extract multi-scale features to address variations in the scale of retinal vessels and combine local and global contextual information to resolve the issue of discontinuity. Xu et al. \cite{xu2020joint} proposed a deep semantics and multi-scaled cross-task aggregation network that leverages the association between retinal vessel segmentation and centerline extraction to jointly improve their performance. Tan et al. \cite{tan2024deep} proposed a W-shaped Deep Matched Filtering method for retinal vessel segmentation, utilizing Anisotropic Perceptive Convolution and Anisotropic Enhancement Module to enhance vascular features and improve orientation and positional accuracy, while suppressing pathological features. Rencently, Amit Bhat et al. \cite{bhati2025dynamic} proposed DyStA-RetNet, a dynamic statistical attention-based lightweight model for retinal vessel segmentation, which employs a shallow CNN-based encoder-decoder architecture. The model incorporates both high-level and low-level information through a partial decoder and utilizes a multi-scale dynamic attention block and a statistical spatial attention block to enhance segmentation performance.

\subsection{CNNs, Vision Transformers and Mamba}
Convolutional neural networks (CNNs) have played a fundamental role in medical image analysis by hierarchically extracting spatial and semantic features, significantly improving segmentation performance. ResNet \cite{he2016deep} alleviates the vanishing gradient problem in deep networks through residual connections; and DenseNet \cite{huang2017densely} enhances feature reuse and gradient flow via densely connected layers. To further improve the modeling of geometric variations and complex structures, Dai et al. \cite{dai2017deformable} proposed Deformable Convolutional Networks, which introduce deformable convolutions and RoI pooling to enhance spatial transformation modeling.

In recent years, Vision Transformers (ViT) have gradually become a popular choice in visual tasks due to their advantages in capturing long-range dependencies and modeling global contextual information. Dosovitskiy et al. \cite{dosovitskiy2020image} proposed the classic ViT model, which demonstrated the superiority of the Transformer-based structure in image classification tasks, breaking the limitations of traditional CNNs. The model divides the image into fixed-size patches and uses these patches as input to capture global contextual information through self-attention mechanisms. Later, the Swin Transformer \cite{liu2021swin} further improved ViT by introducing a hierarchical structure and a window shifting mechanism, optimizing computational efficiency and enhancing the performance of model across various visual tasks. The successful application of ViT and its variants has provided new insights for other visual tasks, such as image segmentation and object detection.

Mamba, a state-of-the-art foundational model, has recently surged in popularity and is now widely used across various visual tasks. VMamba \cite{liu2024vmamba} is the first model to introduce Mamba into vision tasks, addressing the need for efficient network architectures in computer vision. It adapts Mamba into a vision backbone with linear time complexity using Visual State-Space blocks and a 2D Selective Scan (SS2D) module, which captures contextual information across four scanning routes. This structure bridges the gap between ordered 1D scans and the non-sequential nature of 2D vision data. Subsequently, Zhao et al. \cite{zhao2024rs} 
proposed RSMamba (RSM) for remote sensing tasks to address the issue of contextual information loss caused by cropping large images into smaller patches. Since land covers in remote sensing images are distributed in arbitrary directions, RSM incorporates the Omnidirectional Selective Scan Module to globally model the contextual information of images in multiple directions, capturing large-scale spatial features. Mamba is also frequently used in the field of medical image processing. U-Mamba, proposed by Ma \cite{ma2024u}, is a general-purpose network for 3D and 2D biomedical image segmentation. It effectively models long-range dependencies while capturing fine-grained features and achieves linear scalability, making it more efficient compared to the quadratic complexity of Transformer models. Xing et al. \cite{xing2024segmamba} proposed SegMamba, which combines a tri-orientated Mamba module, gated spatial convolution for enhanced spatial feature representation, and a feature-level uncertainty estimation module for better feature reuse, specifically designed for 3D medical image segmentation. However, their approach still has limitations when dealing with elongated medical images, such as those of the retina and blood vessels.
In the work on segmenting elongated targets, He et al. \cite{he2024mamba} proposed CrackMamba, a new module based on the Mamba architecture, specifically designed to enhance the representation of crack features in images. It incorporates an attention mechanism, allowing for more effective capture of the global context of cracks while maintaining computational efficiency. CrackMamba aims to address the limitations of traditional CNNs, which lack global modeling capabilities, and Transformers, which face high computational costs.

\begin{figure*}[t!]
\centering
\includegraphics[width=0.95\textwidth,keepaspectratio]{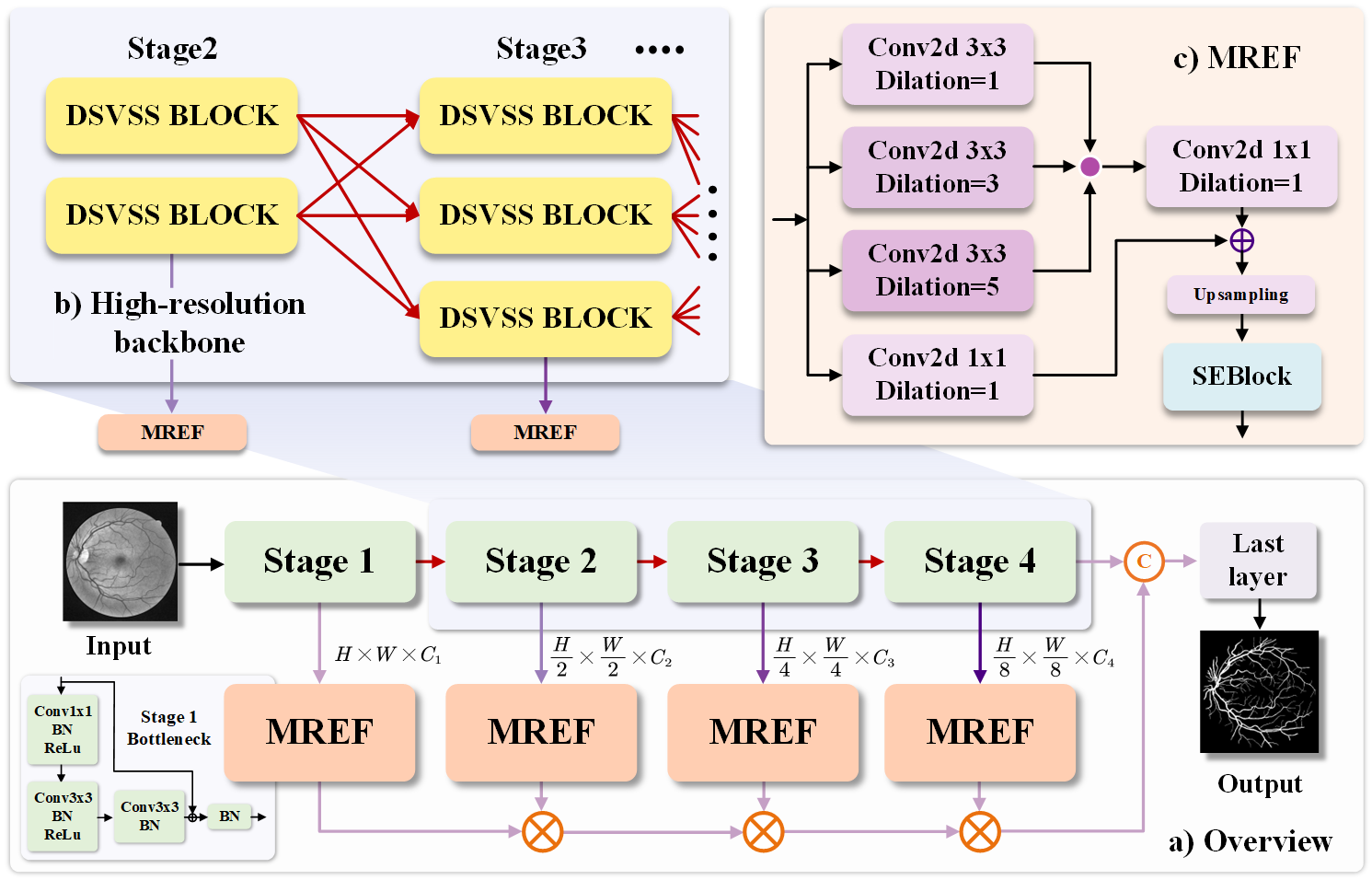}
\caption{\textbf{The overall architecture of the proposed HREFNet.} (a) is overview of HREFNet, where $H$ and $W$ denote the height and width of the input image, respectively, and $C_i$ represents the number of channels. (b) takes Stage 2 and Stage 3 as an example to demonstrate multi-scale fusion details, while this fusion process actually occurs across Stage 2 to Stage 4. (c) represent the Multi-scale Retina Edge Fusion module, designed to enhance edge details and multi-scale feature aggregation}
\label{fig:fig3}
\end{figure*}

\section{METHODOLOGY}
\label{sec:guidelines}
\subsection{Overall Network Architecture}
To address the challenges of retinal vessel segmentation, we propose High-Resolution Edge Fuse Network (HREFNet), a hybrid segmentation framework that integrates multi-scale feature fusion and edge perception mechanisms. As shown in Fig. \ref{fig:fig3} (a), the overall architecture consists of two main components: a high-resolution preserving backbone network and a Multi-scale Retina Edge Fusion (MREF) module. This designed architecture ensures robust and accurate vessel segmentation, particularly in low-contrast regions and complex morphological structures.

As shown in Fig. \ref{fig:fig3} (b), we adopt a multi-stage high-resolution preserving backbone network in our HREFNet to extract multi-scale vessel features while maintaining spatial details. The network progresses through four stages, where the feature streams in each stage not only retain the resolution of the previous stage but also incorporate an additional lower resolution. Previous studies \cite{yun2024shvit} have shown that performing convolution operations on larger feature maps in the early stages facilitates feature extraction. Therefore, our network consists of four stages. In stage 1, we adopt a standard Bottleneck structure composed of 1×1 and 3×3 convolutions with residual connections to extract initial features and facilitate gradient propagation. In stages 2 to 4, we replace standard convolutional blocks with the proposed Dynamic Snake Visual State Space (DSVSS) Block to enhance structural modeling and vessel continuity, which will be described in detail in the following section. Moreover, inspired by HRNet, we perform multi-scale feature fusion through a combination of upsampling and downsampling operations across branches, enabling effective integration of features at different resolutions. This hierarchical representation is essential for accurately segmenting both large vessels and fine capillaries.

Formally, given an input image \( \mathbf{X} \in \mathbb{R}^{H \times W} \), where \( H \) and \( W \) denote the height and width of the image, HREFNet first extracts initial features using a Bottleneck module, producing the output feature map of Stage 1:
\begin{equation}
\begin{aligned}
\mathbf{F}_{0} &= \mathrm{ReLU}(\mathrm{BN}(\mathrm{Conv}_{1\times 1}(\mathbf{X}))), \\
\mathbf{F}_{1} &= \mathrm{ReLU}(\mathrm{BN}(\mathrm{Conv}_{3\times 3}(\mathbf{F}_{0}))), \\
\mathbf{F}_{\mathrm{stage}1} &= \mathrm{ReLU}(\mathrm{BN}(\mathrm{Conv}_{1\times 1}(\mathbf{F}_{1}))) + \mathbf{X}.
\end{aligned}
\end{equation}
where \( \mathrm{Conv}_{k \times k} \) denotes a convolutional layer with kernel size \( k \times k \),  
\( \mathrm{BN} \) and \( \mathrm{ReLU} \) are batch normalization and the ReLU activation function, respectively.  
\( \mathbf{F}_{0} \) and \( \mathbf{F}_{1} \) are the intermediate outputs of the bottleneck, while \( \mathbf{F}_{\mathrm{stage}1} \) denotes the final output.

The feature maps are then processed through three additional stages (Stage 2 to Stage 4), where the output feature maps of the last block of each stage are:
$ F_{\text{stage}2} \in \mathbb{R}^{\frac{H}{2} \times \frac{W}{2} \times C_2}, \quad 
F_{\text{stage}3} \in \mathbb{R}^{\frac{H}{4} \times \frac{W}{4} \times C_3}, \quad
F_{\text{stage}4} \in \mathbb{R}^{\frac{H}{8} \times \frac{W}{8} \times C_4} $. To enhance vessel boundary information, we introduce the MREF module after each backbone stage to fuse multi-scale edge features. Specifically, for each stage output \( \mathbf{F}_{\mathrm{stage}i} \), the MREF module produces an edge-enhanced feature:
\begin{equation}
\mathbf{Q}_i = \mathrm{MREF}(\mathbf{F}_{\mathrm{stage}i}), \quad i \in \{1, 2, 3, 4\},
\end{equation}
where \( \mathrm{MREF}(\cdot) \) denotes the Multi-scale Residual Edge Fusion module, \( \mathbf{F}_{\mathrm{stage}i} \) represents the feature map from Stage \( i \), and \( \mathbf{Q}_i \) is the corresponding edge-enhanced feature output by the MREF module.

These features are subsequently fused through element-wise multiplication to effectively integrate multi-scale edge information:
\begin{equation}
\mathbf{Y}_2 = \mathbf{Q}_1 \odot \mathbf{Q}_2 \odot \mathbf{Q}_3 \odot \mathbf{Q}_4,
\end{equation}
where \( \odot \) denotes the element-wise multiplication operation.

To better capture global context information, we upsample the low-scale outputs from multiple blocks in Stage 4 using bilinear interpolation to restore them to the original input resolution \( H \times W \). These upsampled features are then concatenated along the channel dimension:
\begin{equation}
\mathbf{Y}_1 = \text{Concat} \left( \text{Upsample}(\mathbf{F}_{\mathrm{block},2}), \dots, \text{Upsample}(\mathbf{F}_{\mathrm{block},4}) \right),
\end{equation}
where \( \mathbf{F}_{\mathrm{block},i} \) represents the output of the \( i \)-th block in Stage 4, and \( \text{Upsample}(\cdot) \) denotes the bilinear interpolation upsampling operation.

Finally, \( \mathbf{Y}_1 \) and \( \mathbf{Y}_2 \) are concatenated and processed through the final prediction layer which consists of two convolution layers to fuse the features and generate the final vessel segmentation output:
\begin{equation}
\mathbf{Y}_{\mathrm{seg}} = \text{Lastlayer}(\text{Concat}(\mathbf{Y}_1, \mathbf{Y}_2)),
\end{equation}
where \(\text{Concat}(\cdot)\) denotes channel-wise concatenation, \(\text{Lastlayer}(\cdot)\) represents the final prediction block with two convolutional layers, and \( \mathbf{Y}_{\mathrm{seg}} \in \mathbb{R}^{H \times W} \) is the final segmentation logit with the same spatial resolution as the input image.

The proposed framework effectively combines the local feature extraction capability of CNNs with the long-range dependency modeling capability of Mamba. By integrating the MREF module and high-resolution backbone fusion, the network improves its ability to capture vessel boundaries, especially in low-contrast regions, yielding better segmentation stability and accuracy.

\begin{figure*}[t!]
\centering
\includegraphics[width=0.95\textwidth,keepaspectratio]{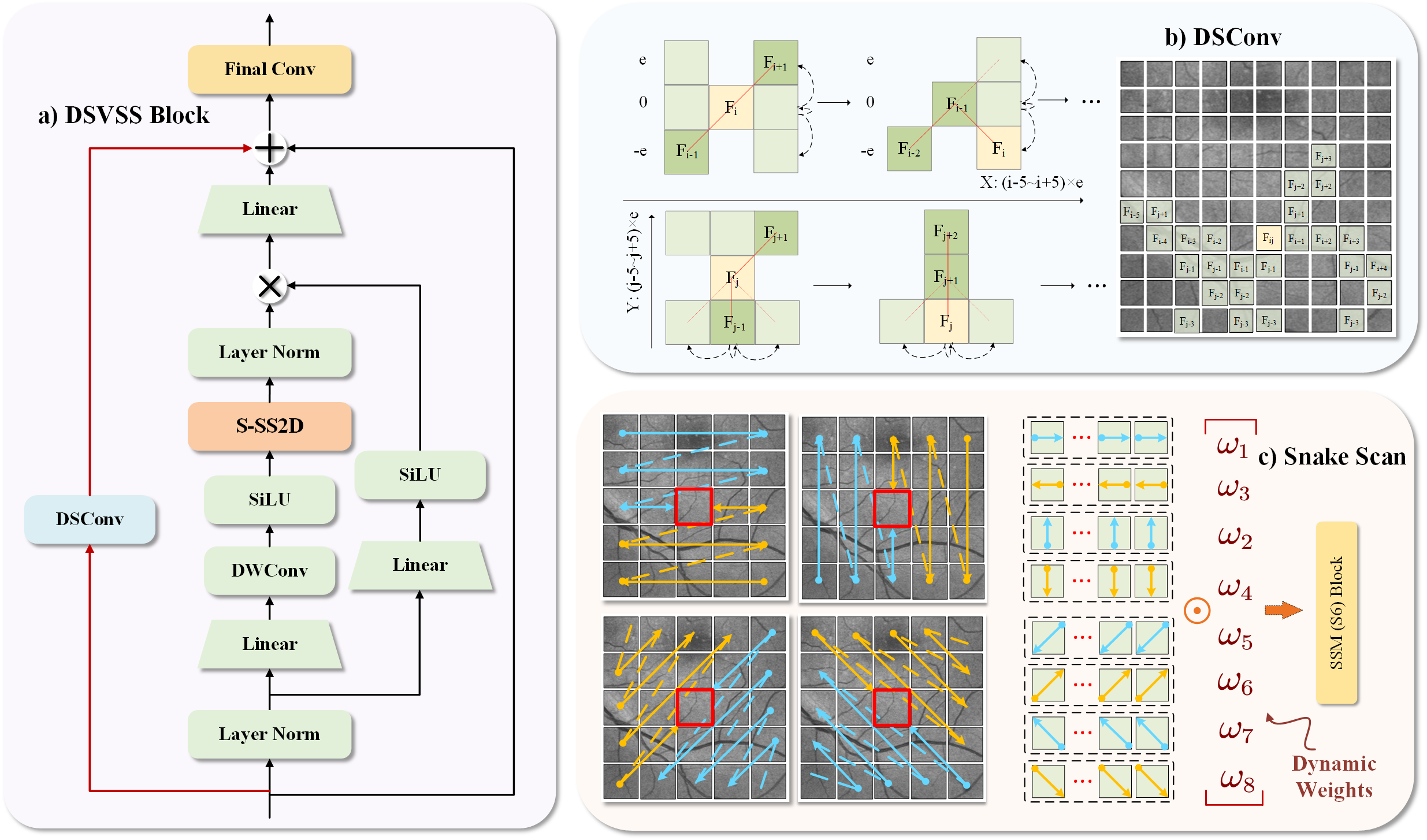}
\caption{
\textbf{Overview of the proposed DSVSS Block.} 
(a) is the structure of the proposed Dynamic Snake Visual State Space Block, which consists of three branches: a dynamic branch with DSConv for adaptive receptive fields, an identity branch, and a DWConv branch with our 2D Snake-Selective Scan (S-SS2D) module for long-range modeling. Outputs from the three branches are fused and refined through a final convolution (Final Conv). (b) presents the Dynamic Snake Convolution, which adjusts the receptive field based on vessel curvature and direction, enhancing feature extraction. (c) depicts the S-SS2D module that performs dynamic weighted scanning across eight directions, enabling effective feature aggregation and improving segmentation of complex vessel structures}
\label{fig:fig4}
\end{figure*}

\subsection{Dynamic Snake Visual State Space (DSVSS) Block}

The Dynamic Snake Visual State Space (DSVSS) block is a key component of the HREFNet architecture, designed to accurately capture complex curvilinear structures in biomedical images. This module integrates high-resolution convolutional processing with dynamic, context-aware state space modeling to enhance vessel segmentation.

To address the limitations of existing dynamic convolutions in capturing global context, we propose the DSVSS Block as an enhanced structural modeling module. Built upon the foundation of Dynamic Snake Convolution (DSConv), which adaptively adjusts the receptive field to better align with vessel curvature and orientation, we extend its capacity by incorporating Mamba-based state space modeling. While DSConv extracts local geometric features along complex vessel paths, its lack of global awareness limits segmentation continuity. The integration of the efficient long-range dependency modeling of Mamba enables the DSVSS Block to capture both fine-grained spatial details and global vessel structures. This joint design significantly improves the robustness of vessel segmentation, particularly for thin, fragmented, or low-contrast vessels, leading to more accurate and stable predictions.

As shown in Fig. \ref{fig:fig4} (a), the DSVSS module consists of three main processing branches. The first branch utilizes DSConv for feature extraction, dynamically adjusting the receptive field to align with the curvature and direction of blood vessels, thereby capturing vessel structures more accurately. The second branch retains the original input features, providing direct contextual information for subsequent feature fusion. The third branch first applies a LayerNorm module for normalization, followed by two sub-branches. The first sub-branch performs nonlinear transformations using a linear layer and a SiLU activation function. The second sub-branch employs a depthwise separable convolution (DWConv) for local feature extraction and further processes the features using the 2D Snake-Selective Scan module (S-SS2D) to capture long-range dependencies and complex spatial information. The outputs from these two sub-branches are fused using a tensor dot product to generate a comprehensive feature representation. After these operations, the features from the three main branches are summed and then aggregated and refined using the final convolution layer (Final Conv). Through this multi-path design, the DSVSS module effectively captures both local spatial information and long-range dependencies, achieving accurate spatial localization. Particularly in scenarios involving thin or fragmented vessels, this module significantly enhances segmentation accuracy.

As shown in Fig. \ref{fig:fig4} (b), the DSVSS module operation dynamically adjusts the receptive field to align with vessel curvature, enhancing feature extraction along intricate vessel paths. The variations along the \( x \)-axis and \( y \)-axis in the receptive field under DSVSS module are formulated as:
\begin{equation}
\begin{aligned}
    F_{i\pm c} &=
    \begin{cases}
        (x_{i+c}, y_{i+c}) = \left( x_i + c,\; y_i + \sum\limits_{i}^{i+c} \left(\Delta y \times e\right) \right), \\
        (x_{i-c}, y_{i-c}) = \left( x_i - c,\; y_i + \sum\limits_{i}^{i-c} \left(\Delta y \times e\right) \right),
    \end{cases} \\
    F_{j\pm c} &=
    \begin{cases}
        (x_{j+c}, y_{j+c}) = \left( x_j + \sum\limits_{j}^{j+c} \left( \Delta x \times e \right),\; y_j + c \right), \\
        (x_{j-c}, y_{j-c}) = \left( x_j + \sum\limits_{j}^{j-c} \left( \Delta x \times e \right),\; y_j - c \right),
    \end{cases}
\end{aligned}
\end{equation}
where, \( \Delta x \) and \( \Delta y \) represent the local variations of the vessel along the \( x \)-axis and \( y \)-axis directions, respectively, while \( c \) denotes the step size adjustment parameter. The hyperparameter \( e \) acts as a curvature modulation factor, dynamically scaling the receptive field adaptation based on local vessel shape. This mechanism allows the receptive field to better align with vessel structures, improving the ability of the model to adapt to varying curvatures and enhancing segmentation accuracy.

The 2D Snake-Selective Scan (S-SS2D) module enhances long-range dependency modeling by performing directional scanning across multiple orientations. Unlike traditional SS2D, which scans in four diagonal directions,  as shown in Fig. \ref{fig:fig4} (c) we extend it to eight directions, including horizontal and vertical scans, to improve feature continuity in complex structures such as vessel bifurcations and thin vessel segments.

We introduce a dynamic weighting mechanism to refine the process of feature extraction. Specifically, each of the eight predefined scanning directions (indexed by \( i = 1, 2, \ldots, 8 \)) is associated with a learnable scalar weight \( w_i \). These weights are normalized using the softmax function to obtain attention coefficients \( \alpha_i \), defined as:
\begin{equation}
    \alpha_i = \frac{\exp(w_i)}{\sum_{j=1}^{8} \exp(w_j)}, \quad i = 1, 2, \ldots, 8
\end{equation}
where, \( \alpha_i \) represents the normalized importance of the \( i \)-th directional feature, ensuring that all directional contributions are adaptively balanced and context-aware. Let \( \mathbf{X}_i \) denote the feature representation extracted along the \( i \)-th direction. The aggregated output feature \( \mathbf{X}' \) is then computed as the weighted sum:
\begin{equation}
    \mathbf{X}' = \sum_{i=1}^{8} \alpha_i \mathbf{X}_i.
\end{equation}
This aggregation strategy is inspired by active contour models and dynamically emphasizes structurally relevant directional cues. As a result, it improves segmentation robustness and preserves spatial and structural consistency across complex regions.

Finally, we integrate it into the DSVSS module, enabling the model to scan along vessel paths. After passing through the S6 block from Mamba, selective information retention is improved, allowing the model to emphasize relevant features while filtering out irrelevant information. This combination creates an efficient segmentation framework that significantly boosts vessel segmentation accuracy, especially in cases where vessel structures exhibit high curvature and variability.

\subsection{Multi-scale Retina Edge Fusion (MREF) Module}

The Multi-scale Retina Edge Fusion (MREF) module is specifically designed to enhance the segmentation of retinal vessels by focusing on edge feature extraction and multi-scale feature fusion. This module integrates adaptive dilation convolutions, channel attention mechanisms, and efficient non-linear activations to capture both fine vessel boundaries and broader contextual information.

Given a feature map \( \mathbf{Z} \in \mathbb{R}^{H \times W \times C} \), MREF employs three dilated convolutions with different dilation rates (\( 1, 3, 5 \)) to effectively capture retinal vessel edge details. Dilated convolutions expand the receptive field without significantly increasing the number of parameters, enabling the extraction of multi-scale features from thin and fragmented vessel structures. The outputs from different dilation rates are then passed through a \( 1 \times 1 \) convolution to refine and unify the extracted features:
\begin{equation}
\begin{aligned}
\mathbf{D}_1 &= \text{ELU}(\text{Conv}_{3\times3}^{\text{d=1}}(\mathbf{Z})), \\
\mathbf{D}_2 &= \text{ELU}(\text{Conv}_{1\times1}^{\text{d=1}}(\text{Conv}_{3\times3}^{\text{d=1}}(\mathbf{Z}))), \\
\mathbf{D}_3 &= \text{ELU}(\text{Conv}_{1\times1}^{\text{d=1}}(\text{Conv}_{3\times3}^{\text{d=3}}(\mathbf{Z}))), \\
\mathbf{D}_4 &= \text{ELU}(\text{Conv}_{1\times1}^{\text{d=1}}(\text{Conv}_{3\times3}^{\text{d=5}}(\mathbf{Z}))).
\end{aligned}
\end{equation}
where, \( \mathbf{D}_1, \mathbf{D}_2, \mathbf{D}_3, \mathbf{D}_4 \) represent different dilation rates in convolutional branches. ELU denotes the Exponential Linear Unit activation function, and \( \text{Conv}_{k \times k}^{\text{d}} \) represents a \( k \times k \) convolution with dilation rate \( d \).

Finally, their outputs are fused via element-wise addition, followed by a residual connection to enhance edge features:
\begin{align}
\mathbf{F}_{\text{edge}} &= \mathbf{D}_1 + \mathbf{D}_2 + \mathbf{D}_3 + \mathbf{D}_4.
\end{align}
To further refine the fused edge features, MREF incorporates the Squeeze-and-Excitation (SE) attention module \cite{hu2018squeeze}, which dynamically recalibrates the importance of channel-wise features. The SE module first applies global average pooling to capture global context, followed by two fully connected layers and a sigmoid activation function to generate channel-wise attention weights:
\begin{equation}
\begin{aligned}
\mathbf{z} &= \text{GlobalAvgPool}(\mathbf{F}_{\text{edge}}), \\
\mathbf{s} &= \sigma(W_2 \mathrm{ReLU}(W_1 \mathbf{z})),
\end{aligned}
\end{equation}
where \( W_1 \) and \( W_2 \) are fully connected layers, \( \mathrm{ReLU} \) denotes the ReLU activation function, and \( \sigma \) is the sigmoid function. The channel-wise attention weights are applied to enhance key edge features while suppressing irrelevant information.
\begin{align}
\mathbf{F}_{\text{enhanced}} = \mathbf{F}_{\text{edge}} \cdot \mathbf{s}.
\end{align}
MREF also incorporates scale-specific processing branches to effectively handle the multi-scale nature of retinal vessels. Depending on the input channel size, the module applies different convolutional layers to ensure consistent output dimensions, thus facilitating effective and efficient cross-scale fusion.

In summary, the MREF module effectively enhances the HREFNet architecture by extracting and fusing multi-scale edge features, integrating adaptive channel attention, and maintaining global contextual consistency. This design significantly improves retinal vessel segmentation performance, particularly in challenging regions with 	thin or low-contrast vessel boundaries.

\section{Experiments setup}
To thoroughly evaluate the performance of the proposed network, a series of extensive experiments are conducted. In this section, we introduce the datasets utilized in our experiments, as well as the evaluation metrics employed to assess the performance of network. A detailed description of the experimental setup and configurations can be found in the subsequent subsections.
\subsection{Datasets}
{To evaluate the performance of the proposed network on the retinal vessel segmentation task, three widely used public fundus image datasets are employed, namely DRIVE \cite{staal2004ridge}, STARE \cite{hoover2000locating}, and CHASE\_DB1 \cite{carballal2018automatic}.
\begin{enumerate}
    \item \textbf{DRIVE (Digital Retinal Images for Vessel Extraction):} The DRIVE dataset consists of 40 retinal fundus images, with 20 images for training and 20 for testing. These images are obtained from a diabetic retinopathy screening program in the Netherlands and have a resolution of 565 × 584 pixels. Each image is manually annotated by an expert, with a second set of annotations provided for reference.

    \item \textbf{STARE (Structured Analysis of the Retina):} The STARE dataset contains 20 retinal fundus images, with 15 images for training and 5 for testing. Initially collected for retinal disease detection, these images have a resolution of 700 × 605 pixels. Each image is accompanied by two sets of manually annotated vessel segmentations from two independent experts. The availability of multiple annotations allows for comparative analysis of segmentation performance.

    \item \textbf{CHASE\_DB1 (Child Heart and Health Study in England):} The CHASE\_DB1 dataset consists of 28 retinal fundus images, with 20 images for training and 8 for testing. These images are collected as part of a child health study in England and primarily depict the retinal fundus of children, with a resolution of 999 × 960 pixels. Each image has been manually annotated by two independent experts, providing a more reliable reference for evaluating model performance.
\end{enumerate}

\subsection{Metrics}

In this study, several evaluation metrics are employed to assess the performance of the segmentation model:

\subsubsection{Dice Coefficient (Dice)}

The Dice coefficient \cite{dice1945measures} is a commonly used metric to evaluate the overlap between two sets. It is calculated as:
\begin{equation}
\mathrm{Dice} = \frac{2 \cdot |\mathit{A} \cap \mathit{B}|}{|\mathit{A}| + |\mathit{B}|} \times 100\%,
\end{equation}
where \( \mathit{A} \) and \( \mathit{B} \) are the sets representing the ground truth and the predicted segmentation, respectively.

\subsubsection{Clustering Dice (\text{clDice})}

The clustering Dice coefficient (clDice) \cite{shit2021cldice} is a topology-aware metric specifically designed to evaluate the structural integrity of thin and elongated anatomical structures, such as blood vessels or neural fibers. Unlike the standard Dice coefficient that measures only region overlap, clDice emphasizes the preservation of connectivity by comparing the skeletons of the predicted and ground truth masks.

Let \( \mathit{S}_\mathit{P} \) and \( \mathit{V}_\mathit{L} \) denote the predicted and ground truth binary masks, respectively. Their corresponding skeletons are \( \mathit{S}_\mathit{P} \) and \( \mathit{V}_\mathit{L} \). The topology precision and topology sensitivity are defined as:
\begin{equation}
\begin{aligned}
\mathrm{Tprec}(\mathit{S}_\mathit{P}, \mathit{V}_\mathit{L}) &= \frac{|\mathit{S}_\mathit{P} \cap \mathit{V}_\mathit{L}|}{|\mathit{S}_\mathit{P}|}, \\
\mathrm{Tsens}(\mathit{S}_\mathit{L}, \mathit{V}_\mathit{P}) &= \frac{|\mathit{S}_\mathit{L} \cap \mathit{V}_\mathit{P}|}{|\mathit{S}_\mathit{L}|}.
\end{aligned}
\end{equation}
The clDice is then computed as the harmonic mean of \(\mathrm{Tprec}\) and \(\mathrm{Tsens}\):
\begin{equation}
\mathrm{clDice}(\mathit{V}_\mathit{P}, \mathit{V}_\mathit{L}) = \frac{2 \cdot \mathrm{Tprec}(\mathit{S}_\mathit{P}, \mathit{V}_\mathit{L}) \cdot \mathrm{Tsens}(\mathit{S}_\mathit{L}, \mathit{V}_\mathit{P})}{\mathrm{Tprec}(\mathit{S}_\mathit{P}, \mathit{V}_\mathit{L}) + \mathrm{Tsens}(\mathit{S}_\mathit{L}, \mathit{V}_\mathit{P})}.
\end{equation}
This formulation ensures that connectivity (precision) and coverage (sensitivity) of the predicted structure relative to the ground truth are optimized together.

\subsubsection{Accuracy (ACC)}

Accuracy measures the proportion of correctly classified pixels (both foreground and background) relative to the total number of pixels:
\begin{equation}
\mathrm{ACC} = \frac{TP + TN}{TP + TN + FP + FN} \times 100\%,
\end{equation}
where \( TP \), \( TN \), \( FP \), and \( FN \) denote the numbers of true positives, true negatives, false positives, and false negatives, respectively.

\subsubsection{Area Under the Curve (AUC)}

The Area Under the Curve (AUC) \cite{bradley1997use} metric evaluates the performance of a binary classifier by computing the area under the Receiver Operating Characteristic (ROC) curve. The AUC is calculated as:
\begin{equation}
\mathrm{AUC} = \int_{0}^{1} \mathrm{TPR} \, \mathrm{d}(\mathrm{FPR}),
\end{equation}
where the true positive rate (TPR) and false positive rate (FPR) are calculated as:
\begin{equation}
\mathrm{TPR} = \frac{TP}{TP + FN}, \quad \mathrm{FPR} = \frac{FP}{FP + TN}.
\end{equation}

\subsubsection{Hausdorff Distance at the 95th Percentile (HD95)}

The Hausdorff Distance at the 95th percentile (HD95) is a robust metric for evaluating spatial discrepancy between predicted and ground truth boundaries. Unlike traditional Hausdorff Distance (HD), HD95 excludes the farthest 5\% of points during computation, reducing the impact of outliers and providing a more stable measure \cite{taha2015metrics}.

\subsection{Details}

Our method is implemented using the PyTorch library on an Nvidia RTX 3090 Ti GPU. To train the proposed segmentation network, the AdamW optimizer is used, with the learning rate passed through command-line arguments, where the default value is set to \(1 \times 10^{-4}\), and the \(\beta\) parameters are set as \((0.9, 0.95)\). The weight initialization method for all convolutional layers is Kaiming initialization.

In the preprocessing stage, the original color fundus images are converted into grayscale images to reduce the influence of color on the segmentation results. Then, image patches of size \(224 \times 224\) are randomly cropped from the fundus images to ensure consistency in input dimensions. During training, the batch size is set to 1, and the model is trained for a total of 300 epochs. The first 200 epochs are pure training, while the last 100 epochs consisted of training with simultaneous validation of the model performance.

For the loss computation, we use the Binary Cross-Entropy (BCE) loss.
In addition, the learning rate is initially set to \(1 \times 10^{-4}\), and the total loss \(L_{\text{all}}\) for each epoch is monitored. The learning rate decayed by a factor of 0.7 once \(L_{\text{all}}\) does not decrease further for two consecutive epochs, until it reaches the minimum value of \(1 \times 10^{-5}\).

\section{Experiment results and analysis }
In this section, we evaluate the overall performance of the proposed network based on experimental results. First, by comparing the segmentation results with state-of-the-art methods and incorporating visualization analysis, we demonstrate the advantages of the proposed network. Then, ablation studies are conducted to verify the effectiveness of the proposed network and each network module. Additionally, further advantages of the proposed modules are highlighted through visualization analysis.

\subsection{Comparison with the state-of-the-art methods}
To evaluate the effectiveness of our proposed HREFNet architecture, we conducted extensive experiments on three widely used datasets: DRIVE, STARE, and CHASE\_DB1. We compared our method against several state-of-the-art approaches covering CNN-based models (e.g., U-Net \cite{ronneberger2015u}, SegResNet \cite{he2016deep}, HRNet \cite{sun2019deep}, and CS-Net \cite{mou2019cs}, DSCNet \cite{qi2023dynamic}), Transformer-based architectures (e.g., HRFormer \cite{yuan2021hrformer}), Mamba-based frameworks (e.g., U-Mamba \cite{ma2024u} and Swin-Umamba \cite{liu2024swin}), and hybrid or task-specific designs (e.g., IMMF-Net \cite{liu2024imff}, and FRNet \cite{ning2024accurate}).

\subsubsection{Quality comparison}
To further demonstrate the effectiveness of our proposed HREFNet, we present a visual comparison of the segmentation results on the DRIVE, STARE, and CHASE\_DB1 datasets, as illustrated in Fig. \ref{fig:fig5}. The red pixels represent false positives, the yellow pixels indicate false negatives, and the green pixels denote true positives. By analyzing the results of different methods, we observe several significant findings.

Firstly, U-Net and HRNet exhibit noticeable false negatives, particularly in low-contrast regions and around thin vessels. This indicates their limited capability in capturing small and intricate vascular structures. HRFormer, leveraging its Transformer-based architecture, partially mitigates this issue; however, its segmentation results still suffer from vessel disconnections and inaccuracies.

DSCNet and U-Mamba demonstrate better vessel connectivity. DSCNet enhances local and global information interaction through dynamic feature aggregation and adaptive feature fusion mechanisms, contributing to improved vessel continuity. U-Mamba, employing visual state space modeling, captures long-range dependencies more effectively, reducing vessel fragmentation. However, both methods still exhibit a higher number of false positives in regions with complex backgrounds.
Swin-UMamba further reduces false negatives by incorporating a hierarchical feature representation. Its multi-scale attention mechanism enhances vessel detection capabilities to some extent. Nevertheless, in certain challenging scenarios, it still struggles to distinguish vessel pixels from non-vessel pixels accurately.

In contrast, our HREFNet demonstrates superior segmentation performance, as evident from the reduced false positives and false negatives. The incorporation of the DSVSS Block and MREF module enables our model to capture multi-scale vessel details and preserve vessel continuity. Additionally, the enhanced edge information provided by the MREF module ensures accurate boundary delineation, reducing the occurrence of fragmented vessels. In the STARE and CHASE\_DB1 datasets, which contain a higher density of thin and overlapping vessels, HREFNet shows greater robustness. The results highlight clearer vessel structures with minimal errors, indicating the capability of our model to adapt to challenging segmentation scenarios.

Overall, the visual results corroborate the quantitative performance gains of HREFNet. By effectively addressing the limitations of existing methods, our proposed model ensures more reliable and accurate retinal vessel segmentation, contributing to improved diagnostic support in clinical applications.

\begin{figure*}[t!]
\centering
\includegraphics[width=0.95\textwidth,keepaspectratio]{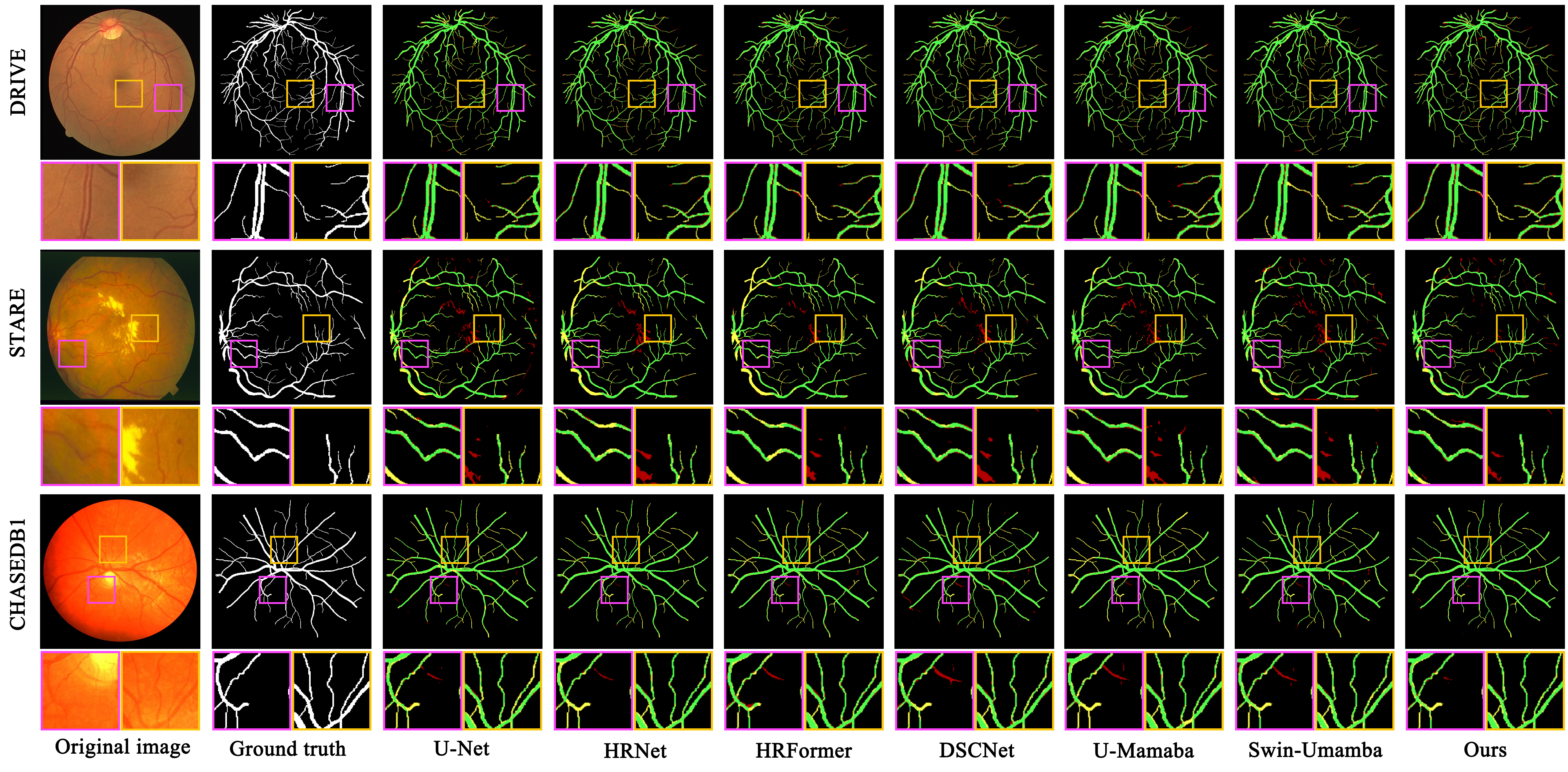}
\caption{
\textbf{Typical segmentation results of different methods on three classical datasets.} DRIVE (top), STARE (middle), CHASE\_DB1 (bottom). The red pixel indicates false positives, the yellow pixel indicates false negatives, and the green pixels represent the true positives
}
\label{fig:fig5}
\end{figure*}

\begin{table}[ht]
    \centering
    \caption{Comparison with state-of-the-art methods on DRIVE. The best scores are \textbf{highlighted} in bold, and the second best score is \underline{underlined}}
    \label{tab:results1}
    \begin{tabular}{@{\extracolsep{\fill}}lccccc@{}}
        \toprule
        \textbf{Method} & \textbf{Dice (\%)} & \textbf{clDice (\%)} & \textbf{ACC (\%)} & \textbf{AUC (\%)} & \textbf{HD95 (pixel)} \\
        \midrule
        U-Net \cite{ronneberger2015u} & $79.54 \pm 1.91$ & $78.44 \pm 4.22$ & $96.58 \pm 0.31$ & $97.71 \pm 0.60$ & $8.74 \pm 3.08$ \\
        SegResNet \cite{he2016deep} & $80.52 \pm 1.74$ & $80.94 \pm 4.12$ & $96.65 \pm 0.28$ & $97.99 \pm 0.56$ & $6.52 \pm 2.81$ \\
        HRNet \cite{sun2019deep} & $81.47 \pm 1.67$ & $81.10 \pm 4.03$ & $96.90 \pm 0.28$ & $98.10 \pm 0.68$ & $6.98 \pm 3.45$ \\
        CS-Net \cite{mou2019cs} & $80.52 \pm 2.04$ & $79.84 \pm 4.36$ & $96.74 \pm 0.29$ & $98.03 \pm 0.51$ & $8.65 \pm 3.87$ \\
        HRFormer \cite{yuan2021hrformer} & $81.83 \pm 2.13$ & $81.90 \pm 4.12$ & $\mathbf{97.00 \pm 0.34}$ & $\underline{98.40 \pm 0.53}$ & $\underline{4.88 \pm 2.68}$ \\
        DSCNet \cite{qi2023dynamic} & $81.42 \pm 1.67$ & $81.71 \pm 4.09$ & $96.78 \pm 0.26$ & $98.19 \pm 0.51$ & $6.35 \pm 2.87$ \\
        U-Mamba \cite{ma2024u} & $81.69 \pm 1.73$ & $81.70 \pm 4.28$ & $\underline{96.90 \pm 0.30}$ & $98.30 \pm 0.55$ & $5.37 \pm 2.87$ \\
        IMMF-Net \cite{liu2024imff} & $\underline{81.85 \pm 1.57}$ & $\underline{82.38 \pm 4.07}$ & $96.83 \pm 0.26$ & $98.21 \pm 0.59$ & $5.24 \pm 2.73$ \\
        FRNet \cite{ning2024accurate} & $80.51 \pm 2.40$ & $80.01 \pm 4.53$ & $96.77 \pm 0.34$ & $98.34 \pm 0.41$ & $8.23 \pm 3.65$ \\
        Swin-Umamba \cite{liu2024swin} & $\underline{81.85 \pm 1.55}$ & $82.30 \pm 3.72$ & $96.80 \pm 0.25$ & $98.30 \pm 0.56$ & $\mathbf{4.68 \pm 2.16}$ \\
        HREFNet(ours) & $\mathbf{82.14 \pm 1.53}$ & $\mathbf{82.40 \pm 4.07}$ & $\underline{96.90 \pm 0.24}$ & $\mathbf{98.56 \pm 0.39}$ & $5.45 \pm 3.02$ \\
        \bottomrule
    \end{tabular}
\end{table}

\begin{table}[ht]
    \centering
    \caption{Comparison with state-of-the-art methods on STARE. The best scores are \textbf{highlighted} in bold, and the second best score is \underline{underlined}}
    \label{tab:results2}
    \begin{tabular}{@{\extracolsep{\fill}}lccccc@{}}
        \toprule
        \textbf{Method} & \textbf{Dice (\%)} & \textbf{clDice (\%)} & \textbf{ACC (\%)} & \textbf{AUC (\%)} & \textbf{HD95(pixel)} \\
        \midrule
         U-Net \cite{ronneberger2015u} & $70.10 \pm 2.15$ & $72.28 \pm 4.12$ & $95.83 \pm 0.30$ & $96.50 \pm 0.64$ & $22.20 \pm 3.88$ \\
          SegResNet \cite{he2016deep} & $74.95 \pm 1.61$ & $78.77 \pm 3.03$ & $96.47 \pm 0.51$ & $97.33 \pm 0.48$ & $17.14 \pm 3.51$ \\
        HRNet \cite{sun2019deep} & $71.96 \pm 8.52$ & $77.40 \pm 5.52$ & $96.50 \pm 0.93$ & $97.50 \pm 1.05$ & $21.40 \pm 8.24$ \\
        CS-Net \cite{mou2019cs} & $63.21 \pm 13.82$ & $67.64 \pm 10.15$ & $95.80 \pm 1.06$ & $96.30 \pm 1.39$ & $28.93 \pm 15.62$ \\
        HRFormer \cite{yuan2021hrformer} & $69.60 \pm 10.50$ & $74.30 \pm 8.27$ & $96.30 \pm 0.98$ & $96.50 \pm 1.54$ & $25.10 \pm 13.30$ \\
        DSCNet \cite{qi2023dynamic} & $75.69 \pm 2.66$ & $80.10 \pm 2.76$ & $\underline{96.60 \pm 0.51}$ & $97.70 \pm 0.57$ & $\mathbf{16.30 \pm 4.32}$ \\
        U-Mamba \cite{ma2024u} & $73.54 \pm 9.39$ & $78.60 \pm 6.60$ & $\mathbf{96.70 \pm 0.92}$ & $\mathbf{98.10 \pm 0.66}$ & $18.90 \pm 11.10$ \\
        IMMF-Net \cite{liu2024imff} & $72.54 \pm 9.21$ & $77.80 \pm 6.09$ & $96.57 \pm 0.85$ & $97.67 \pm 0.83$ & $20.18 \pm 10.16$ \\
        FRNet \cite{ning2024accurate} & $62.99 \pm 0.15$ & $66.66 \pm 12.24$ & $95.56 \pm 0.96$ & $94.74 \pm 2.69$ & $35.96 \pm 18.31$ \\
        Swin-Umamba \cite{liu2024swin} & $\underline{76.08 \pm 2.87}$ & $\underline{80.20 \pm 2.65}$ & $\underline{96.60 \pm 0.89}$ & $97.90 \pm 0.69$ & $\underline{16.70 \pm 3.50}$ \\
        HREFNet(ours) & $\mathbf{76.29 \pm 3.94}$ & $\mathbf{80.30 \pm 3.65}$ & $\mathbf{96.70 \pm 0.78}$ & $\underline{98.05 \pm 0.69}$ & $17.10 \pm 5.83$ \\
        \bottomrule
    \end{tabular}
\end{table}

\begin{table}[ht]
    \centering
    \caption{Comparison with state-of-the-art methods on CHASE\_DB1. The best scores are \textbf{highlighted} in bold, and the second best score is \underline{underlined}}
    \label{tab:results3}
    \begin{tabular}{@{\extracolsep{\fill}}lccccc@{}}
        \toprule
        \textbf{Method} & \textbf{Dice (\%)} & \textbf{clDice (\%)} & \textbf{ACC (\%)} & \textbf{AUC (\%)} & \textbf{HD95(pixel)} \\
        \midrule
        U-Net \cite{ronneberger2015u} & $76.56 \pm 1.71$ & $78.10 \pm 2.48$ & $96.94 \pm 0.37$ & $97.86 \pm 0.65$ & $16.45 \pm 3.05$ \\
        SegResNet \cite{he2016deep} & $78.61 \pm 1.76$ & $79.79 \pm 2.63$ & $97.31 \pm 0.37$ & $98.21 \pm 0.54$ & $16.89 \pm 3.06$ \\
        HRNet \cite{sun2019deep} & $79.80 \pm 2.73$ & $80.90 \pm 3.55$ & $\underline{97.50 \pm 0.44}$ & $98.70 \pm 0.44$ & $24.20 \pm 11.70$ \\
        CS-Net \cite{mou2019cs} & $72.52 \pm 3.84$ & $69.33 \pm 4.62$ & $96.95 \pm 0.47$ & $97.02 \pm 0.86$ & $40.82 \pm 12.14$ \\
        HRFormer \cite{yuan2021hrformer} & $79.45 \pm 2.18$ & $80.76 \pm 3.14$ & $97.41 \pm 0.42$ & $98.32 \pm 0.47$ & $20.88 \pm 7.73$ \\
        DSCNet \cite{qi2023dynamic} & $79.89 \pm 1.79$ & $82.10 \pm 2.89$ & $97.40 \pm 0.32$ & $98.70 \pm 0.35$ & $13.40 \pm 4.53$ \\
        U-Mamba \cite{ma2024u} & $75.91 \pm 2.81$ & $74.47 \pm 3.21$ & $97.19 \pm 0.44$ & $97.34 \pm 1.04$ & $39.76 \pm 10.63$ \\
        IMMF-Net \cite{liu2024imff} & $77.77 \pm 2.24$ & $77.00 \pm 2.62$ & $97.30 \pm 0.46$ & $98.40 \pm 0.41$ & $31.20 \pm 7.50$ \\
        FRNet \cite{ning2024accurate} & $71.01 \pm 3.50$ & $66.20 \pm 4.75$ & $96.74 \pm 0.42$ & $96.90 \pm 1.13$ & $45.34 \pm 14.38$ \\
        Swin-Umamba \cite{liu2024swin} & $\underline{80.20 \pm 1.70}$ & $\underline{82.40 \pm 2.49}$ & $\mathbf{97.60 \pm 0.33}$ & $\mathbf{98.80 \pm 0.40}$ & $\underline{12.53 \pm 3.64}$ \\
        HREFNet(ours) & $\mathbf{80.46 \pm 1.57}$ & $\mathbf{82.93 \pm 2.50}$ & $97.36 \pm 0.37$ & $\underline{98.78 \pm 0.35}$ & $\mathbf{10.34 \pm 3.17}$ \\
        \bottomrule
    \end{tabular}
\end{table}

\subsubsection{Quantitative comparison}
To quantitatively assess the experimental results, we conducted a statistical comparison using key performance metrics including Dice, clDice, ACC, AUC, and HD95 across three datasets: DRIVE, STARE, and CHASE\_DB1. Our proposed method is evaluated against state-of-the-art approaches to ensure performance validation.

Table \ref{tab:results1} shows the performance comparison on the DRIVE dataset. Our method achieved the highest Dice score of 82.14\% and clDice of 82.40\%, surpassing all competing methods. In particular, our model demonstrated the highest AUC of 98.56\%, indicating superior segmentation quality. While HRFormer achieved a slightly higher ACC of 97.00\%, our method maintained a competitive ACC of 96.90\%, showcasing a well-balanced performance. Additionally, our HD95 score of 5.45 further highlights the robustness of our segmentation.

For the STARE dataset, as presented in Table \ref{tab:results2}, our model attained the best Dice score of 76.29\% and a clDice of 80.30\%, outperforming other approaches. Although the AUC value of our method (98.05\%) is slightly lower than that of Umamba (98.10\%), it remained highly competitive. Notably, our ACC score of 96.70\% is also the highest, reflecting the effectiveness of our model in accurately segmenting vessels. The HD95 value of 17.10 demonstrated satisfactory boundary agreement.

On the CHASE\_DB1 dataset (Table \ref{tab:results3}), our proposed method continued to show superior performance. It achieved the highest Dice score of 80.46\% and a clDice of 82.93\%, exhibiting an evident improvement over previous methods. Our AUC score of 98.78\% is only marginally lower than that of Swin-Umamba. However, our method yielded a remarkable HD95 score of 10.34, the lowest among all models, indicating accurate boundary segmentation.

In summary, our proposed model consistently outperformed other state-of-the-art methods in most metrics across the DRIVE, STARE, and CHASE\_DB1 datasets. The superior AUC, Dice, and HD95 values indicate the effectiveness and robustness of our segmentation network for retinal vessel analysis.

\subsection{Ablation study}
\subsubsection{Ablation Study on DSVSS Block}
To evaluate the effectiveness of the innovative components within the proposed Dynamic Snake Visual State Space (DSVSS) block, we conducted an ablation study on the DRIVE dataset. The results are summarized in Table \ref{tab:ablation1}, where we use metrics such as the Dice coefficient, clDice, accuracy (ACC), AUC, and HD95 to assess the performance of the model. The baseline model consists of an HRNet with the standard Visual State Space Block (VSSBlock). We incorporated the DSConv and S-SS2D components to analyze their individual and combined contributions to the model.

Fig. \ref{fig:fig6} provides a visual comparison of segmentation results for different configurations. Notably, as shown in columns (e) and (f), adding S-SS2D and DSConv contributes to finer vessel boundary segmentation. Compared to baseline results in column (d), clearer and more continuous vessel structures are observed. Additionally, integrating components in column (g) refines vessel segmentation, demonstrating enhanced vessel connectivity and suppressed background noise.

The DSConv component introduces dynamic state context normalization, which refines feature representations by enhancing vessel boundary clarity. As seen in Table \ref{tab:ablation1}, incorporating DSConv into the baseline model improved the Dice score from 81.64\% to 81.86\% and increased the clDice from 81.95\% to 82.11\%. The AUC also saw a notable improvement from 98.15\% to 98.30\%, demonstrating enhanced segmentation accuracy. The Hausdorff Distance (HD95) is reduced from 5.60 to 5.47, indicating better structural continuity in vessel predictions.

The S-SS2D component further enhances vessel boundary segmentation by leveraging a dynamic 2D state space for curvilinear structure modeling. Adding S-SS2D to the baseline model yielded consistent improvements, with the Dice score increasing to 81.75\% and the clDice reaching 82.03\%. Additionally, the AUC rose to 98.28\%, and the HD95 slightly increased to 5.79. The results suggest that S-SS2D effectively captures vessel boundary information, especially in complex regions. This is further validated in Fig. \ref{fig:fig6}, where clearer and smoother vessel contours can be observed in column (e) compared to the baseline.

When both the DSConv and S-SS2D components are integrated into the baseline model, significant performance gains are observed. Specifically, the Dice score reached 81.92\%, a 0.28\% increase over the baseline. The clDice improved to 82.20\%, with a gain of 0.25\%. Notably, the AUC peaked at 98.60\%, a 0.45\% improvement, highlighting the enhanced ability of the model to distinguish vessels from the background. Moreover, accuracy increased to 96.90\%, reflecting a 0.35\% rise and superior vessel segmentation quality. Although the HD95 is slightly reduced to 5.56, the overall segmentation boundary consistency improved. The visual results in Fig. \ref{fig:fig6} column (g) further demonstrate the effectiveness of the combined approach, showing sharper vessel boundaries and fewer false positives.

\begin{figure*}[t!]
\centering
\includegraphics[width=0.95\textwidth,keepaspectratio]{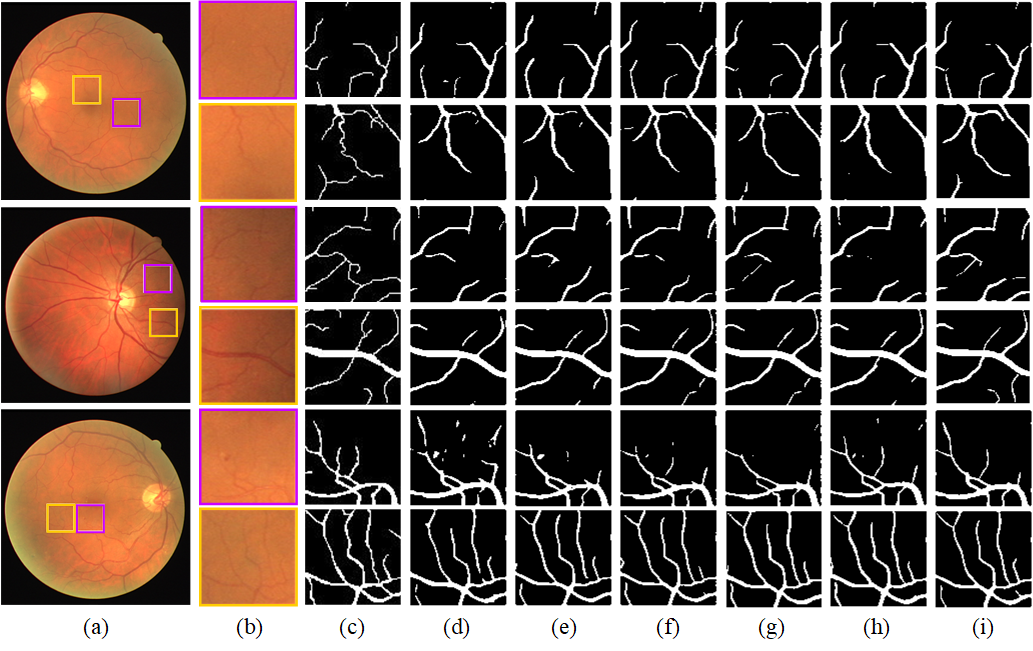}
\caption{
\textbf{Some typical visual results for different methods in our ablation study on DRIVE dataset.} (a) Original image, (b) detailed view, (c) ground truth, (d) baseline, (e) baseline+S-SS2d, (f) baseline+DSConv, (g) baseline+DSVSS Block, (h) baseline+MREF, (i) baseline+DSVSS Block+MREF(Ours)
}
\label{fig:fig6}
\end{figure*}
\begin{table*}[ht] 
\centering
\caption{Component-wise Ablation Study of the Proposed DSVSS Block on the DRIVE Dataset. Checkmarks ($\checkmark$) indicate the inclusion of each component. The best scores are \textbf{highlighted} in bold, and the second best score is \underline{underlined}}
\label{tab:ablation1}
\resizebox{\textwidth}{!}{  
\begin{tabular}{@{}lccccS[table-format=2.2]S[table-format=2.2]S[table-format=3.2]S[table-format=2.2]S[table-format=3.2]@{}}
\toprule
\textbf{Model} & \textbf{VSSblock} & \textbf{DSConv} & \textbf{S-SS2D} & \textbf{Dice (\%)} & \textbf{clDice (\%)} & \textbf{ACC (\%)} & \textbf{AUC(\%)} & \textbf{HD95(pixel)} \\
\midrule
Baseline & $\checkmark$ & $\times$ & $\times$ & 81.64 & 81.95 & 96.55 & 98.15 & 5.60 \\
Baseline + DSConv   & $\checkmark$ & $\checkmark$ & $\times$ & \underline{81.86} & \underline{82.11} & \underline{96.74} & \underline{ 98.30} & \phantom{1}\textbf{5.47} \\
Baseline + S-SS2D & $\checkmark$ & $\times$ & $\checkmark$ & 81.75 & 82.03 & 96.67 & 98.28 & 5.79 \\
Baseline + DSConv + S-SS2D  & $\checkmark$ & $\checkmark$ & $\checkmark$ & \textbf{81.92} & \textbf{82.20} & \textbf{96.90} & \phantom{1}\textbf{98.60} &  \underline{ 5.56} \\
\bottomrule
\end{tabular}
}
\vspace{0.2cm}
\end{table*}

\begin{table*}[ht] 
\centering
\caption{Component-wise Ablation Study of the Proposed HREFNet on the DRIVE Dataset. Checkmarks ($\checkmark$) indicate the inclusion of each component. The best scores are \textbf{highlighted} in bold, and the second best score is \underline{underlined}}
\label{tab:ablation2}
\resizebox{\textwidth}{!}{  
\begin{tabular}{@{}lccccS[table-format=2.2]S[table-format=2.2]S[table-format=3.2]S[table-format=2.2]S[table-format=3.2]@{}}
\toprule
\textbf{Model} & \textbf{VSSblock} & \textbf{DSVSS Block} & \textbf{MREF} & \textbf{Dice (\%)} & \textbf{clDice (\%)} & \textbf{ACC (\%)} & \textbf{AUC(\%)} & \textbf{HD95(pixel)} \\
\midrule
Baseline & $\checkmark$ & $\times$ & $\times$ & 81.64 & 81.95 & 96.55 & 98.15 & 5.60 \\
Baseline + DSVSS Block   & $\times$ & $\checkmark$ & $\times$ & \underline{81.92} & \underline{82.20} & \underline{96.90} & \phantom{1}\textbf{98.60} & 5.56 \\
Baseline + MREF & $\checkmark$ & $\times$ & $\checkmark$ & 81.84 & 82.05 & 96.78 & 98.49 & \underline{ 5.52} \\
Baseline + DSVSS Block + MREF  & $\times$ & $\checkmark$ & $\checkmark$ & \textbf{82.14} & \textbf{82.40} & \textbf{96.90} & \underline{ 98.56} &  \phantom{1}\textbf{5.45} \\
\bottomrule
\end{tabular}
}
\vspace{0.2cm}
\end{table*}

\subsubsection{Ablation Study on HREFNet}
To further investigate the effectiveness of different components within the proposed HREFNet, we conduct a comprehensive ablation study on the DRIVE dataset. Table \ref{tab:ablation2} presents the results, where the inclusion of the DSVSS Block, and MREF is examined.

Fig. \ref{fig:fig6} illustrates the visual improvements achieved by adding each component. Specifically, column (h) shows the effect of adding the MREF module, which enhances vessel edge clarity and reduces false positives. When both DSVSS Block and MREF are applied column (i), the segmentation results exhibit finer vessel structures and improved boundary continuity, highlighting the synergistic effect of both modules.

When incorporating the DSVSS Block instead of the VSSblock, the model shows notable improvements in all metrics, achieving a Dice score of 81.92\% and a clDice score of 82.20\%. The AUC reaches 98.60\%, demonstrating the effectiveness of dynamic state space modeling in capturing complex vessel structures, and significantly improving segmentation accuracy.

By adding the MREF module to the baseline, the Dice score increases to 81.84\%, while the clDice score rises to 82.05\%. The improvement in vessel boundary clarity is evident, as the HD95 reduces to 5.52, indicating a more accurate representation of vessel edges. The multi-scale edge fusion mechanism of the MREF module enhances boundary segmentation and provides robustness in low-contrast regions.

When both the DSVSS Block and MREF module are integrated, HREFNet achieves the best overall performance. The Dice score reaches 82.14\%, representing a 0.50\% increase over the baseline. The clDice score increases to 82.40\%, a 0.45\% improvement. Accuracy (ACC) rises to 96.90\%, reflecting a 0.35\% gain. Moreover, HD95 is further reduced to 5.45, showing a decrease of 0.15 compared to the baseline. The combined effect of dynamic state space modeling and multi-scale edge fusion significantly enhances vessel segmentation accuracy and robustness.

These results validate the effectiveness of each proposed module, demonstrating that the combination of DSVSS Block and MREF leads to substantial improvements in both vessel structure identification and boundary refinement. The proposed HREFNet effectively captures fine-grained vessel details while ensuring robust segmentation performance across various vessel sizes and challenging regions.

\subsubsection{Ablation Study on Stage Depth}

To investigate the effect of network depth on segmentation performance, we conduct an ablation study by designing three HREFNet variants with different numbers of  stages: HREFNet-tiny (2 stages), HREFNet-middle (3 stages), and HREFNet-large (4 stages). All three variants share the same backbone components and modules, differing only in their architectural depth.
Table~\ref{tab:ablation_stages} presents the quantitative results on the DRIVE dataset. HREFNet-tiny achieved a Dice score of 81.75\% and a clDice score of 81.90\%, while HREFNet-middle obtained slightly lower scores of 81.64\% and 81.72\%, respectively. HREFNet-large outperformed both, with a Dice score of 82.14\% and a clDice score of 82.40\%. In addition, HREFNet-large achieved the lowest HD95 value of 5.45 pixels, indicating superior boundary precision.
\begin{table}[b] 
\centering
\caption{Performance comparison of HREFNet variants with different stage depths on the DRIVE dataset. The best scores are \textbf{highlighted} in bold, and the second best score is \underline{underlined}}
\label{tab:ablation_stages}
\begin{tabularx}{\columnwidth}{p{3cm} X X X X X} 
\toprule
\textbf{Model} & \textbf{FLOPs} & \textbf{Params} & \textbf{Dice} & \textbf{clDice} & \textbf{HD95} \\
 & (G) & (M) & (\%) & (\%) & (pixel) \\
\midrule
HREFNet-tiny & 150.27 & 23.91 & \underline{81.75} & \underline{81.90} & \underline{5.59} \\
HREFNet-middle & 304.76 & 49.62 & 81.64 & 81.72 & 5.73 \\
HREFNet-large & 446.89 & 92.69 & \textbf{82.14} & \textbf{82.40} & \textbf{5.45} \\
\bottomrule
\end{tabularx}
\end{table}

\begin{figure}[!t]
\centering
\includegraphics[width=0.95\columnwidth,keepaspectratio]{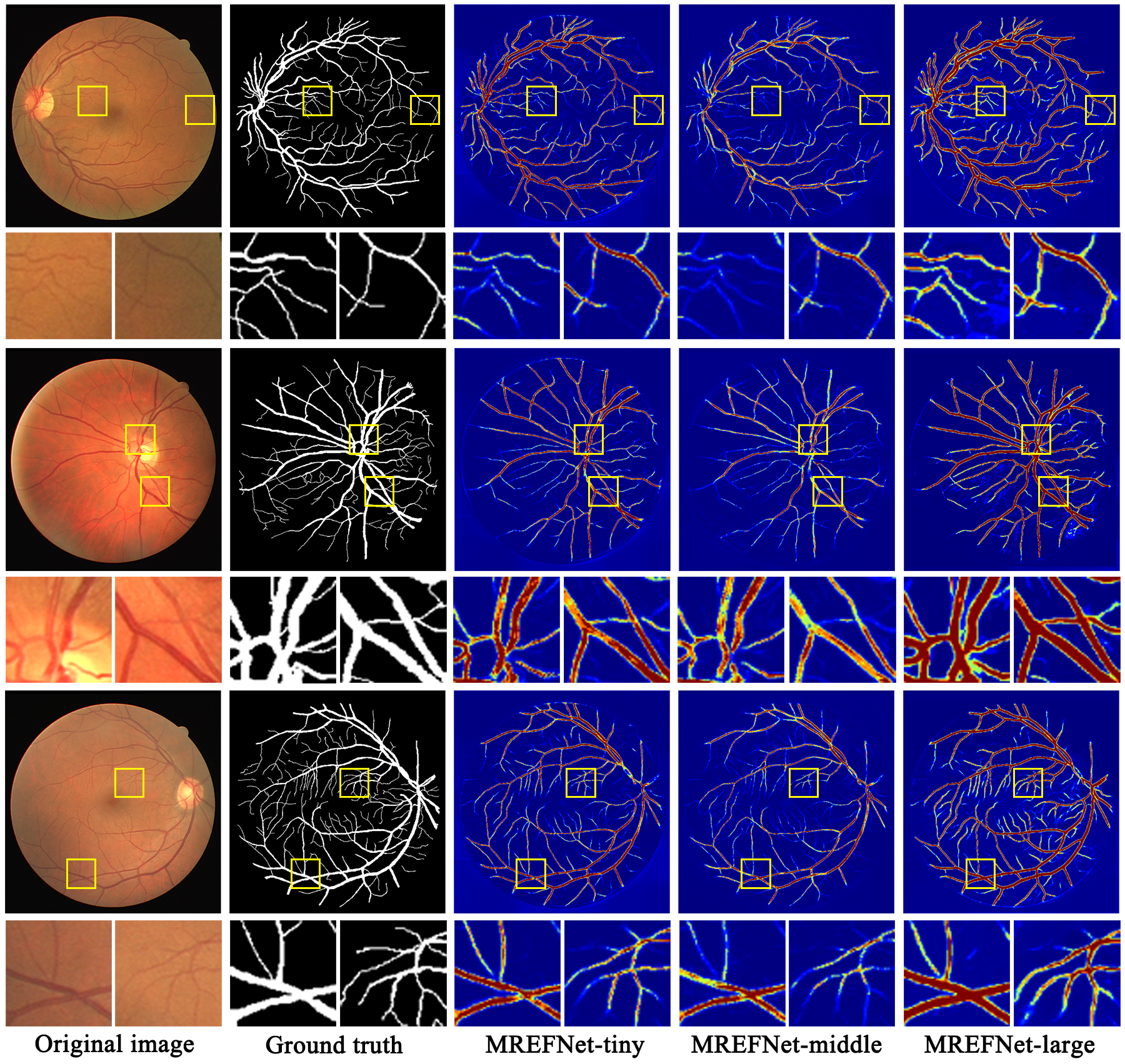}
\caption{\textbf{Grad-CAM visualization of HREFNet variants.} Qualitative comparison of segmentation results from HREFNet-tiny, HREFNet-middle, and HREFNet-large using Grad-CAM heatmaps. HREFNet-large exhibits better vessel continuity, finer details, and clearer boundaries}
\label{fig:gradcam}
\end{figure}

From Table~\ref{tab:ablation_stages}, we observe that HREFNet-large consistently achieves the best performance across all metrics, demonstrating the benefits of deeper architectures in capturing fine-grained vessel structures. Interestingly, HREFNet-tiny slightly outperforms HREFNet-middle in both Dice and clDice scores, suggesting that increasing the number of stages does not always lead to improved performance.

This observation can be attributed to the trade-off between representational capacity and optimization difficulty. HREFNet-tiny, with fewer stages, preserves more spatial details and benefits from easier gradient propagation. In contrast, HREFNet-middle introduces additional complexity without a proportional gain in feature representation, resulting in marginal performance degradation.

The superior performance of HREFNet-large indicates that increasing network depth enhances hierarchical representation learning and improves modeling of long-range dependencies and small vessel structures. The lower HD95 confirms boundary localization. The improvements observed with HREFNet-large highlight the importance of multi-scale edge perception in our architecture: deeper networks capture fine vessel contours at smaller scales effectively, thus improving segmentation accuracy.

To more clearly demonstrate the visual differences, Fig.~\ref{fig:gradcam} presents qualitative segmentation results along with Grad-CAM \cite{selvaraju2017grad} heatmaps for HREFNet-tiny, HREFNet-middle, and HREFNet-large.

As shown in Fig.~\ref{fig:gradcam}, HREFNet-large produces the most complete vessel structures, particularly in regions with low contrast and thin vessels. In contrast, HREFNet-tiny and HREFNet-middle tend to generate fragmented segments or less distinct boundaries, indicating their limitations in capturing fine structural details.

These findings confirm that HREFNet-large offers the optimal trade-off between architectural complexity and segmentation performance. By leveraging a deeper stage design, it significantly enhances hierarchical representation and edge perception, leading to superior retinal vessel segmentation.

\section{Conclusion}
In this paper, we propose a novel hybrid framework, HREFNet, for retinal vessel segmentation. By integrating the Multi-scale Retina Edge Fusion (MREF) module and the Dynamic Snake Visual State Space (DSVSS) module into a high-resolution preserving backbone, HREFNet effectively enhances boundary representation and accurately captures fine vessel structures. These innovations enable the model to achieve accurate segmentation even under challenging conditions such as thin, fragmented, or low-contrast vessels. HREFNet has been extensively evaluated on several standard retinal vessel segmentation benchmarks, consistently outperforming existing state-of-the-art methods. The MREF module improves edge localization, while the DSVSS module enhances long-range dependency modeling and structural detail representation. Ablation studies further confirm the effectiveness of each component and highlight the synergy between multi-scale feature extraction and dynamic modeling. HREFNet demonstrates promising potential in medical image analysis, where its high-quality segmentation results can provide reliable structural information for downstream tasks such as disease detection, diagnosis, and progression assessment. Future work will focus on reducing computational complexity and exploring the generalization capability of HREFNet across other biomedical image segmentation tasks. Overall, HREFNet offers a robust and efficient solution for retinal vessel segmentation and contributes valuable insights to the development of advanced medical image analysis techniques.

\bibliography{sn-bibliography}

\authorbio{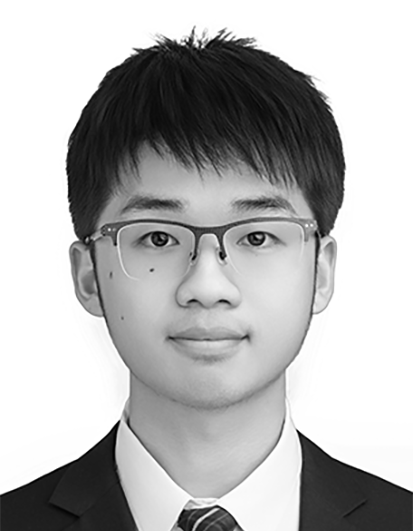}{Yihao Ouyang}{
is currently pursuing the B.S. degree with the School of Artificial Intelligence, Hebei University of Technology, Tianjin, China. His research interests include medical image processing, artificial intelligence, barely supervised learning, computer vision, deep learning, medical image segmentation, label-efficient learning, self-training strategies, uncertainty modeling, and the application of neural networks in clinical decision support systems.
}

\authorbio{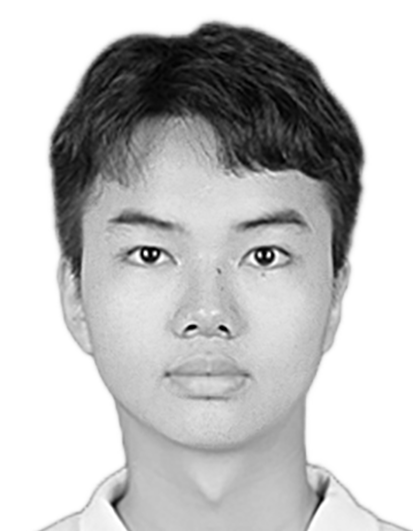}{Xunheng Kuang}{
is currently pursuing the B.S. degree with the School of Artificial Intelligence, Hebei University of Technology, Tianjin, China. His research interests include medical image processing, artificial intelligence, deep learning, and barely supervised learning, with a focus on developing efficient algorithms for medical data analysis and exploring advanced learning techniques under limited supervision.
}

\authorbio{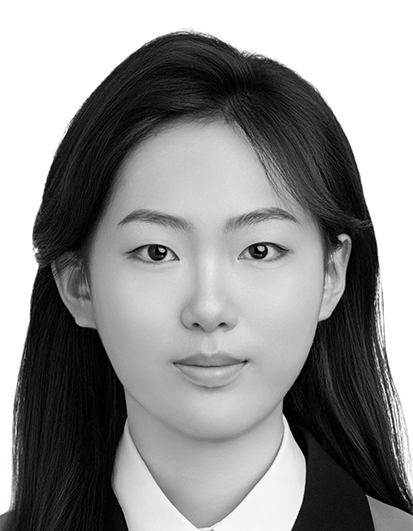}{Mengjia Xiong}{
is currently pursuing the B.S. degree with the School of Artificial Intelligence, Hebei University of Technology, Tianjin, China. His research interests include biomedical signal analysis, neural network optimization, multi-modal learning, medical image reconstruction, intelligent diagnostic systems, as well as the development of algorithms for medical data analysis and the application of deep learning techniques to improve the accuracy and efficiency of clinical decision-making.
}

\authorbio{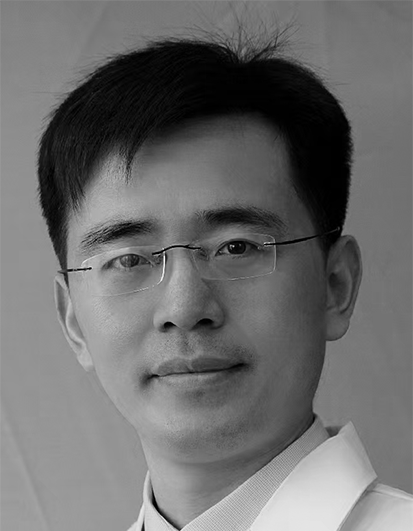}{Zhida Wang}{
holds a Doctor of Medicine degree and completed his postdoctoral research in clinical medicine. He is currently an Associate Professor and Chief Physician at Chu Hsien-I Memorial Hospital and the Tianjin Institute of Endocrinology, Tianjin Medical University. He also serves as a member of the Precision Medicine Committee of the Tianjin Medical Association, the Epidemiology Branch of the Tianjin Cardiology Society, and the 13th Tianjin Youth Federation. His primary research interests focus on diabetes and its related complications.
}

\authorbio{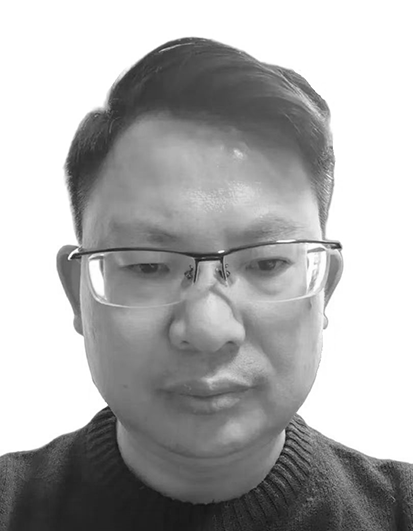}{Yuanquan Wang}{
received his M.S. and Ph.D. degrees from Nanjing University of Science and Technology in 1998 and 2004, respectively. In December 2006, he completed his postdoctoral research at the School of Computer Science, Beijing Institute of Technology. He is currently a professor at the School of Artificial Intelligence, Hebei University of Technology. His research interests include image processing, computer vision, deep learning, and their applications in medical artificial intelligence and object perception.
}
\end{document}